\documentclass{INTERSPEECH2023}

\usepackage{comment}
\usepackage[nolist,nohyperlinks]{acronym}
\usepackage{url}
\usepackage{booktabs}
\usepackage{amsmath}    
\usepackage{textcomp}
\usepackage{graphicx}
\usepackage{subfig}
\usepackage{booktabs}
\usepackage[colorinlistoftodos,prependcaption]{todonotes}
\usepackage{xcolor}
\usepackage{pifont}
\usepackage{tabularx,colortbl}
\usepackage{cite}
\usepackage{amssymb}
\usepackage[]{hyperref}

\acrodef{BWE}{Bandwidth Extension}
\acrodef{GAN}{Generative Adversarial Network}
\acrodef{SSL}{Self-Supervised Learning}
\acrodef{CNN}{Convolutional Neural Network}
\acrodef{FFN}{Point-wise Feed-Forward Network}
\acrodef{MLS}{Multi-Lingual Librispeech}
\acrodef{BERT}{Bidirectional Encoder Representations from Transformers}
\acrodef{HuBERT}{Hidden-unit BERT}
\acrodef{ASR}{Automatic Speech Recognition}
\acrodef{ASV}{Automatic Speaker Verification}
\acrodef{AEE}{Acoustic Environment Embedding}
\acrodef{DINO}{Self-DIstillation with NO labels}
\acrodef{CGAN}{Conditional GAN}
\acrodef{CycleGAN}{Cycle-consistent GAN}
\acrodef{MHA}{Multi-Head Attention}
\acrodef{FiLM}{Feature-wise Linear Modulation}
\acrodef{DNN}{Deep Neural Network}
\acrodef{TSV}{Telephony Speaker Verification}
\acrodef{EER}{Equal Error Rate}
\acrodef{DFL}{Deep feature Loss}
\acrodef{ASR}{Automatic Speech Recognition}
\acrodef{OOD}{out-of-domain}
\acrodef{LMFB}{Log-Mel FilterBank}
\acrodef{AAM}{Additive Angular Margin}
\acrodef{SRE}{Speaker Recognition Evaluation}
\acrodef{minDCF}{minimum Decision Cost Function}
\acrodef{AAM}{Additive Angular Margin}
\acrodef{PLDA}{Probabilistic Linear Discriminant Analysis}

\graphicspath{{images/}}

\def\arrvline{\hfil\kern\arraycolsep\vline\kern-\arraycolsep\hfilneg}

\interspeechcameraready

\title{Self-FiLM: Conditioning GANs with self-supervised representations for bandwidth extension based speaker recognition}
\name{Saurabh Kataria$^{1,2}$, Jes\'us Villalba$^{1,2}$, Laureano Moro-Vel\'azquez$^{1}$, \\Thomas Thebaud$^{1}$, Najim Dehak$^{1,2}$}
\address{
$^{1}$Center for Language and Speech Processing, $^{2}$Human Language Technology Center of Excellence\\
Johns Hopkins University, Baltimore, MD, USA
  }
\email{\{skatari1,jvillal7,laureano,tthebau1,ndehak3\}@jhu.edu}

\begin{document}

\maketitle

\begin{abstract}
Speech super-resolution/Bandwidth Extension (BWE) can improve downstream tasks like Automatic Speaker Verification (ASV).
We introduce a simple novel technique called Self-FiLM to inject self-supervision into existing BWE models via Feature-wise Linear Modulation.
We hypothesize that such information captures domain/environment information, which can give zero-shot generalization.
Self-FiLM Conditional GAN (CGAN) gives 18\% relative improvement in Equal Error Rate and 8.5\% in minimum Decision Cost Function using state-of-the-art ASV system on SRE21 test.
We further by 1) deep feature loss from time-domain models and 2) re-training of data2vec 2.0 models on naturalistic wideband (VoxCeleb) and telephone data (SRE Superset etc.).
Lastly, we integrate self-supervision with CycleGAN to present a completely unsupervised solution that matches the semi-supervised performance.
\end{abstract}
\noindent\textbf{Index Terms}: Self-supervision, FiLM conditioning, conditional GAN, super-resolution, CycleGAN, data2vec 2.0

\section{Introduction}
\label{sec:intro}
Deep learning has incredibly advanced speech applications like source separation, speech enhancement, and \ac{BWE}~\cite{kataria2022joint}.
Such inverse problems involve representation learning and feature mapping between domains.
The rise of \ac{SSL} representations has called for a joint investigation of SSL and BWE (our application of interest).
Conditional variants of generative models like \ac{GAN}~\cite{kataria2022joint}, and diffusion model have shown great promise~\cite{han2022nu} for BWE.
We aim to bridge the gap between SSL and deep generative models by learning to condition \ac{GAN} with SSL representations.

SoundFilter~\cite{gfeller2021one} is a one-shot source separation model which uses a short target utterance as conditioning using Feature-wise Linear Modulation (FiLM)~\cite{perez2018film}.
In \cite{song2023exploring}, authors use a pre-trained WavLM~\cite{chen2022wavlm} SSL model (with additional fine-tuning step) for another input to speech enhancement network.
TUNet~\cite{nguyen2022tunet} uses temporal FiLM-based UNet architecture for BWE and simple self-supervision losses but does not use any conditioning information.
\cite{sukhadia2022channel} pursues mixed-bandwidth \ac{ASR} by doing channel-aware pre-training in a HuBERT~\cite{hsu2021hubert}-inspired \ac{SSL} model.
\cite{pasad2021layer} showed that bottom layers are more suitable for speaker verification while top layers are suited well for \ac{ASR}.

We develop SSL-conditioned BWE models to assist telephony ASV (downstream task)~\cite{kataria2022time}, which prior work still needs to address.
Considering SSL representation as a proxy for \ac{AEE}, we explore the zero-shot adaptation capability of our system during inference using AEE information.
We choose to condition the hidden layers of the BWE model and not provide SSL embedding sequence as an additional input (for e.g., to the first layer) to avoid re-designing the BWE-ASV pipeline.
Also, we require continuous target prediction for BWE, and mapping only SSL embeddings to desired temporal output will be challenging.
With our proposed scheme Self-FiLM, we first establish the utility of various pre-trained SSL models with CGAN.
We also visualize the conditioning information by speaker, language, and domain label as explored similarly in \cite{pasad2021layer}.
Building on an efficient SSL model such as data2vec 2.0, we explore in-domain training as done in Robust wav2vec 2.0~\cite{hsu2021robust}.
We also study Self-FiLM with \ac{DFL} for speaker preservation and CycleGAN for unsupervised learning, which is unavailable in prior work.

In our contributions, we
1) provide the first study to explore a combination of bandwidth extension and self-supervised models in depth (via Self-FiLM),
2) visualize SSL embeddings via FiLM per domain label speaker recognition downstream task,
3) explore re-training of data2vec 2.0 on naturalistic mixed-bandwidth data with deep feature loss-equipped bandwidth extension, 
4) demonstrated compatibility of CycleGAN with data2vec 2.0-based Self-FiLM.

\begin{figure*}[htbp]
    \centering
    \includegraphics[width=0.90\textwidth]{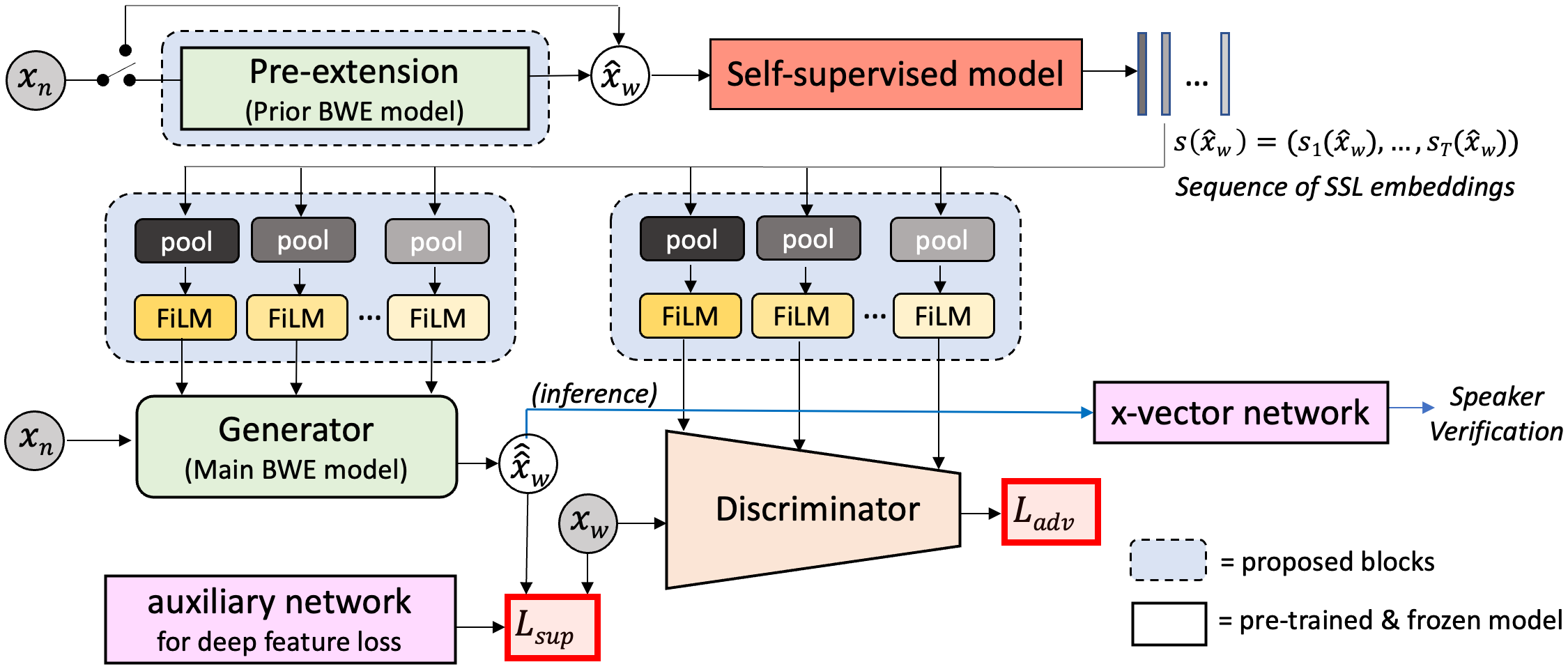}
    \caption{Illustration of the Self-FiLM framework for Conditional GAN bandwidth extension.
    The generator and discriminator receives information from self-supervised model after pooling and FiLM operation.
    An optional \emph{pre-extension} block is present before the SSL model.
    An auxiliary network is used for deriving speaker preserving (deep feature) loss.
    }
    \label{fig:selffilm}
    \vspace{-5mm}
\end{figure*}

\section{Background}
\label{sec:back}
\subsection{Self-supervised learning models}
\label{sec:selfsup}
We extract 256-D frame-level representations from primarily pre-trained small/BASE version of 16KHz SSL audio models.
since they capture low as well as high-level information~\cite{pasad2021layer}.
\\
\underline{\textbf{wav2vec 2.0}}~\cite{baevski2020wav2vec}:
This model is trained with a contrastive loss defined over jointly learned quantization of latent representations.
In the BASE version (95M parameters), the feature encoder consists of seven convolutional blocks.
The context/transformer network has 12 layers, 768 model dimensions, 3072 \ac{FFN} inner dimension, eight attention heads, and relative positional encoding.
The training data is Librispeech~\cite{panayotov2015librispeech} which consists of 960h of read speech.
To test multi-lingual generalization on our test sets, we also experiment with \emph{Robust Large wav2vec 2.0} (0.3B parameters), which is trained on Libri-light, CommonVoice, Switchboard, and Fisher~\cite{hsu2021robust}.
\\
\underline{\textbf{XLSR-53}}~\cite{conneau2020unsupervised}:
This is another multi-lingual counterpart of wav2vec 2.0 (BASE) trained with 53 languages from 
CommonVoice~\cite{ardila2019common} (read speech), BABEL~\cite{gales2014speech} (conversational telephone speech), and \ac{MLS}~\cite{pratap2020mls} (read speech from audiobooks).
\\
\underline{\textbf{WavLM}}~\cite{chen2022wavlm}:
This model jointly accomplishes masked speech target prediction (like HuBERT~\cite{hsu2021hubert}) and denoising.
The denoising capability makes WavLM more conducive to non-ASR tasks.
Through simulation, noisy input samples are created.
The BASE model (94.7M parameters) is trained on Librispeech and has similar architecture to wav2vec 2.0.
\\
\underline{\textbf{Data2vec 2.0}}~\cite{baevski2022data2vec}:
This BASE model (93.8M parameters) is derived from a non-contrastive student-teacher formulation.
Using unmasked input, the teacher model produces the output.
Using the masked input, the student model tries to predict the activations produced in various layers of the teacher model.
Teacher follows a running exponential average of the student.

\subsection{Embedding pooling methods}
\label{sec:pool}
\underline{\textbf{mean}, \textbf{mean+std}}:
Here, \emph{mean} refers to simple average pooling across time dimension, while \emph{mean+std} (\emph{statistics pooling}) refers to the concatenation of mean and standard deviation.
\\
\underline{\textbf{LDE}}~\cite{cai2018exploring}:
In Learnable Dictionary Encoding (LDE), we assume that the representations $(\mathbf{s_1}, \dots, \mathbf{s_T})$ are in $C$ clusters.
We learn a dictionary with the centers as $\{\mu_c\}$, where $1\leq c \leq C$.
Soft cluster assignment $w_{t,c}$ and final embedding $e$ are
\begin{equation}
    w_{t,c} = \frac{\exp(-||\mathbf{s_t} - \mu_c||^2)}{\sum_{c=1}^{C}\exp(-||\mathbf{s_t} - \mu_c||^2)},
\end{equation}
\begin{equation}
    e_c = \frac{\sum_{t=1}^{T} w_{t,c} (\mathbf{s_t} - \mu_c)}{\sum_{t=1}^{T} w_{t,c}}, e = [e_1^T, \dots, e_C^T]^T.
\end{equation}
\\
\underline{\textbf{ScaleAtt}}:
We use a modified form of scaled dot-product attention~\cite{vaswani2017attention}, which we term as ScaleAtt.
We use multiple heads ($H=4$) like in the \ac{MHA}~\cite{india2019self} formulation.
For a single head, attention outputs are
\begin{equation}
\begin{split}
    \text{ScaleAtt} = \text{softmax}(\frac{qK^T}{\sqrt{d_k}})V, 
    K = f_k(\mathbf{x}), V = f_v(\mathbf{x}).
\end{split}
\end{equation}
$q$ is a learnable query vector (per each head), which makes the formulation non self-attentive.
$K$ and $V$ are key and value matrices obtained through $f_k$ and $f_v$ projection linear layers (output dimensions are $d_v=d_v=256$).
$H$ parallel attention modules are utilized, and outputs are concatenated to form final $d_{\text{model}}$ dimension output, where $d_{\text{model}}=Hd_v$.

\begin{table*}[htbp]
\centering
\caption{\label{tab:explore}
Effect of different pooling methods and SSL model in Self-FiLM on ASV metrics: EER/minDCF-formatted (lower the better).
We use pre-trained SSL
and do not investigate \emph{pre-extension} block and \emph{deep feature loss} (Fig.~\ref{fig:selffilm}).
* denotes identical systems.
}
\resizebox{0.98\textwidth}{!}{
\begin{tabular}{@{}lccccc@{}}
\toprule
 \textbf{CGAN BWE type}  & \textbf{CGAN Sup loss} & \textbf{G+D params} & \textbf{SRE16-YUE-eval40} & \textbf{SRE-CTS-superset-dev} & \textbf{SRE21-audio-eval} \\
    \hline
    No BWE & - & - & 7.12 / 0.376 & 5.36 / 0.216 & 17.12 / 0.644 \\
    BWE without self-FiLM & 0.0048 & 1.7 & 5.68 / 0.332 & 4.08 / 0.180 & 15.81 / 0.618 \\  
    \hline
    \textbf{\textit{Pooling type (SSL fixed as wav2vec 2.0)}} & & & & & \\
    mean & 0.0049 & 5.7 & 5.41 / 0.306 & 4.05 / 0.180 & 15.11 / 0.610 \\ 
    mean+std & 0.0051 & 7.6 & 7.09 / 0.335 & 4.01 / 0.178 & 13.84 / \textbf{0.586} \\     
    LDE & 0.0052 & 18.2 & \textbf{5.07} / \textbf{0.298} & \textbf{3.86} / \textbf{0.175} & 14.33 / 0.593 \\ 
    ScaleAtt (*) & 0.0052 & 39.3 & 5.22 / 0.303 & 4.03 / 0.181 & 14.09 / 0.591 \\  
    \hline
    \textbf{\textit{SSL type (Pooling type fixed as ScaleAtt)}} & & & & & \\
    wav2vec 2.0 (*) & 0.0052 & 39.3 & 5.22 / 0.303 & 4.03 / 0.181 & 14.09 / 0.591 \\ 
    Robust Large wav2vec 2.0 & \textbf{0.0035} & 49.2 & 5.44 / 0.307 & 4.00 / 0.177 & 15.42 / 0.614 \\ 
    XLSR-53 & \textbf{0.0035} & 49.2 & 5.52 / 0.306 & 3.96 / 0.177 & 15.32 / 0.611 \\ 
    WavLM & 0.0036 & 39.3 & 5.51 / 0.302 & 3.98 / 0.177 & 15.05 / 0.603 \\ 
    data2vec 2.0 & \textbf{0.0035} & 39.3 & 5.48 / 0.308 & \textbf{3.94} / \textbf{0.175} & 16.07 / 0.621 \\ 
\bottomrule
\end{tabular}
}
\vspace{-5mm}
\end{table*}

\subsection{Speaker embedding networks}
\label{sec:spkemb}
\underline{\textbf{Light-ResNet34}}~\cite{villalba2020advances}:
LResNet~\cite{villalba2020advances} is a smaller ResNet-based x-vector architecture with 256-D embedding, 80-D \ac{LMFB} input features, and 5.6M parameters.
It has four residual blocks whose outputs are used for DFL.
\\
\underline{\textbf{RawNet3}}~\cite{jung2022pushing}:
We utilize this 16.2M parameter time-domain model since it is compatible with self-supervised techniques~\cite{jung2022pushing}.
The first layer is a parametric analytic filter-bank (256 filters) followed by three residual backbone blocks (1024 filters each). 
We use the outputs of all convolutional blocks before the pooling layer for DFL.

\subsection{Generative Adversarial Networks}
\label{sec:gan}
For BWE model, we use Generative Adversarial Networks.
We primarily focus on supervised GANs (Conditional GAN). 
Supervised \ac{CGAN}~\cite{mirza2014conditional} learns to sample from conditional distribution ($p_{A|B}$) where $A$ and $B$ are two domains ($p_A$, $p_B$ resp.).
Generator $\mathcal{G}_{A\rightarrow B}$ generates fake sample while discriminator $D_B$ distinguishes between real and fake via
\begin{equation}
\label{eq:cgan}
    \max_{\mathcal{G}_{A\rightarrow B}} \min_{D_B} \mathcal{L}_{\text{CGAN}}, \hspace{0.5em} \text{where} \hspace{0.5em}
        \mathcal{L}_{\text{CGAN}} = \mathcal{L}_{\text{adv}} + \lambda_{\text{sup}}\mathcal{L}_{\text{sup}}.
\end{equation}
$\mathcal{L}_{\text{adv}}$ and $\mathcal{L}_{\text{sup}}$ (weighted by $\lambda_{\text{sup}}$) are adversarial and supervised loss respectively:
\begin{align}
    \label{eq:adv}
    \mathcal{L}_{\text{adv}}(\textbf{a},\textbf{b}) = &\mathbb{E}_{\textbf{a},\textbf{b} \sim p_{A,B}}
    [(\mathcal{D}(\textbf{b}))^2 + (1 - \mathcal{D}(\mathcal{G}(\textbf{a}))^2
    ],\\
    \mathcal{L}_{\text{sup}}(\textbf{a},\textbf{b}) = &\mathbb{E}_{\textbf{a}\sim p_A, \textbf{b}\sim p_B} [\lVert \textbf{b} - \mathcal{G}_{A\rightarrow B}(\textbf{a})\rVert_1].
\end{align}
Here, $\mathcal{L}_{\text{adv}}$ is based on Least-squares GAN~\cite{mao2017least}.
$\mathbf{a}$ and $\mathbf{b}$ are real paired samples from domains A and B respectively.

CycleGAN is an unpaired model using two tied CGANs:
\begin{align}
\label{eq:cyclegan}
    &\max_{\mathcal{G}_{A\rightarrow B}, \mathcal{G}_{B\rightarrow A}} \min_{\mathcal{D}_{A}, \mathcal{D}_{B}} \mathcal{L}_{\text{cyc-GAN}} \hspace{0.5em} \text{, where} \hspace{2em}\\
        \mathcal{L}_{\text{cyc-GAN}} &= \mathcal{L}_{\text{adv},A\rightarrow B} + \mathcal{L}_{\text{adv},B\rightarrow A} +
        \lambda_{\text{cyc}}\mathcal{L}_{\text{cyc}} + \lambda_{\text{id}}\mathcal{L}_{\text{id}}.
\end{align}
Adversarial losses
are defined like in Eq.~\ref{eq:adv}.
$\mathbf{a}$ and $\mathbf{b}$ are real unpaired samples.
$\lambda_{\text{cyc}}$ and $\lambda_{\text{id}}$ are the weights for cycle and identity loss which are used for semantic consistency and regularization, respectively:
\begin{align}
    \mathcal{L}_{\text{cyc}} &= \mathbb{E}_{\mathbf{a}\sim p_{A},\mathbf{b}\sim p_{B}} [\lVert\mathbf{a} - \mathcal{G}_{B\rightarrow A}(\mathcal{G}_{A\rightarrow B}(\mathbf{a}))\rVert_1]\notag\\
    &+ \mathbb{E}_{\mathbf{a}\sim p_{A},\mathbf{b}\sim p_{B}} [\lVert\mathbf{b} - \mathcal{G}_{A\rightarrow B}(\mathcal{G}_{B\rightarrow A}(\mathbf{b}))\rVert_1],\\
    \mathcal{L}_{\text{id}} &= \mathbb{E}_{\mathbf{a}\sim p_{A}}[\lVert\mathbf{a} - \mathcal{G}_{B\rightarrow A}(\mathbf{a})\rVert_1] \notag \\
    &+ \mathbb{E}_{\mathbf{b}\sim p_{B}}[\lVert\mathbf{b} - \mathcal{G}_{A\rightarrow B}(\mathbf{b})\rVert_1].
\end{align}

\subsection{Generator and Discriminator architectures}
For Generator, we use Conv-TasNet~\cite{luo2019conv} -- a popular time-domain source separation model.
Based on 1-D \ac{CNN}, it consists of \emph{encoder}, \emph{separator}, and \emph{decoder}.
The \emph{separator} has eight CNN layers that compute a mask.
The number of input and output channels for \emph{encoder} and \emph{decoder} is one.
The number of parameters is 1.6M, the kernel size is 16, the stride is 8, and channels increase from 128 to 1024 exponentially with a dilation of 2.
For discriminator, we use Parallel WaveGAN (0.16M parameters). 
It is a 10-layer 1-D CNN with a kernel of 3, channels of 80, and linearly increasing dilation from the second to the ninth layer (from one to eight).

\subsection{Feature-wise Linear Modulation}
\label{sec:film}
For data index $i$ and channel index $c$, \ac{FiLM} operation/layer~\cite{perez2018film} adaptively modulates the activations $F_{i,c}$ of a \ac{DNN} with a conditioning vector $s_i$.
It uses learned linear layers ($f_c$, $h_c$, $g_c$) per channel.
After introducing a FiLM strength hyper-parameter $\alpha$,
\begin{equation}
    \begin{split}
        \gamma_{i,c} &= f_c(g_c(s_i)), \beta_{i,c} = h_c(g_c(s_i)), \\
       \text{FiLM}(F_{i,c}) &= F_{i,c} + \alpha(\gamma_{i,c}F_{i,c} + \beta_{i,c} - F_{i,c}).
    \end{split}
    \label{eq:film}
\end{equation}
Output dimension of $f_c$ and $h_c$ equals $F_{\cdot,c}$ channel dimension.
Dimension of $s_{\cdot}$ depends on pooling function, so we introduce $g_c$ linear layer with output dimension of 256 for standardization.


\begin{figure*}[htbp]
    \centering
    \subfloat[\centering
    Clustering by speaker identity
    ]{{\includegraphics[width=5.00cm]{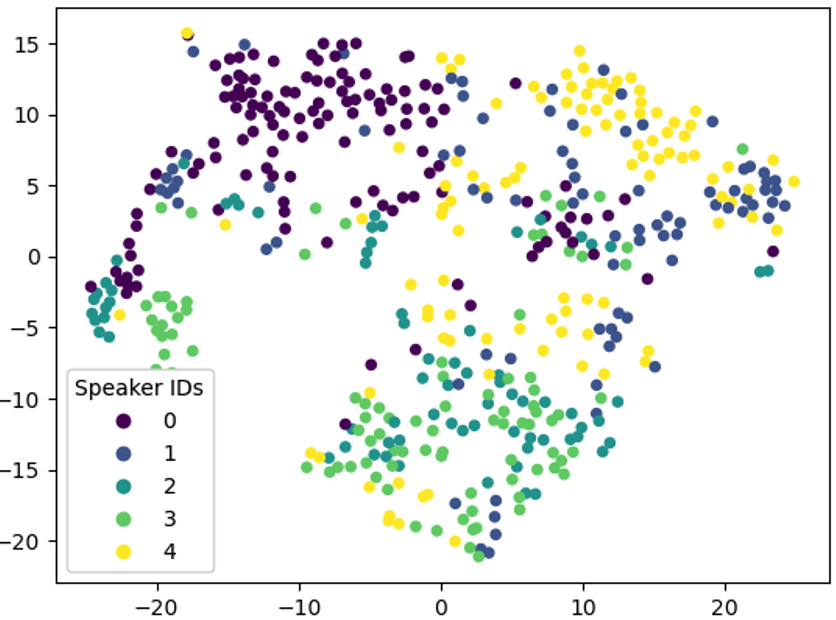}}}%
    \qquad
    \subfloat[\centering
    Clustering by language identity
    ]{{\includegraphics[width=5.00cm]{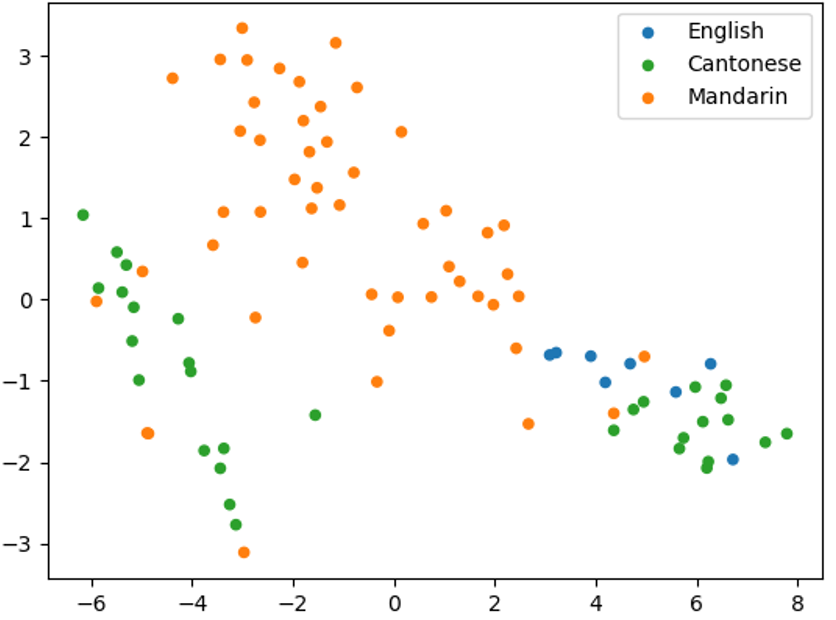}}}%
    \qquad
    \subfloat[\centering
    Clustering by domain: CTS vs AFV
    ]{{\includegraphics[width=5.00cm]{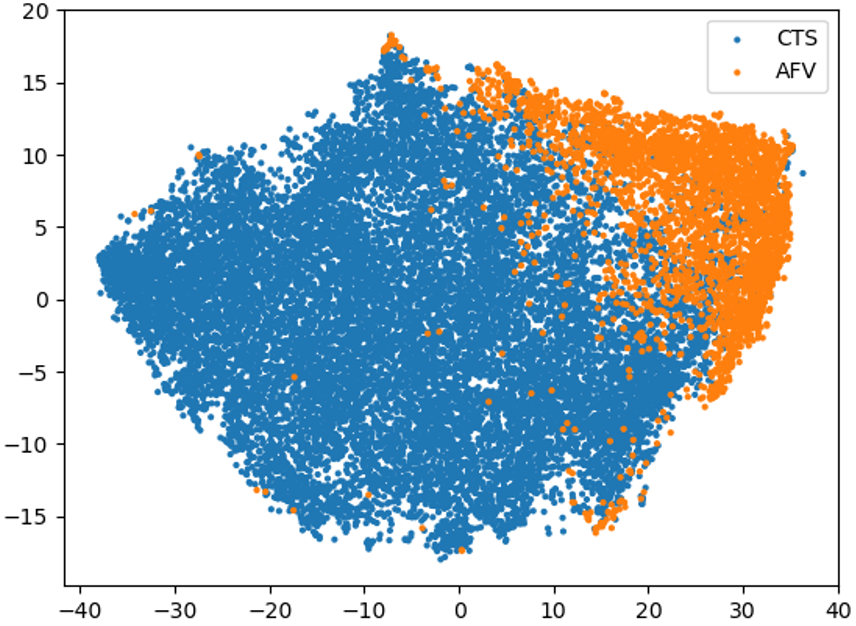}}}%
    \caption{t-SNE visualization of FiLM activations from the first layer of CGAN generator.
    }
    \label{fig:analysis}
    \vspace{-5mm}
\end{figure*}

\section{Self-FiLM}
\label{sec:selffilm}
In Self-FiLM, we condition the layers of a \ac{DNN} with the self-supervised representations of the input signal \emph{itself}.
It provides alternate richer \emph{view} of the signal.
Our flexible framework allows clean integration of SSL models with existing pipelines and avoids system re-design.
The test signal may benefit from SSL conditioning (proxy for acoustic environment embedding) for zero-shot generalization during inference.
In Fig.~\ref{fig:selffilm}, we demonstrate Self-FiLM on Conditional GAN (both generator and discriminator).
Narrowband signal $x_n$ is input to an optional preliminary BWE model (\emph{pre-extension}) for improved compatibility with the (usually) wideband-trained SSL model.
The self-supervised representations are pooled (Sec.~\ref{sec:pool}) and FiLM-ed to various layers of G and D.
We use the modified FiLM proposed in Eq.~\ref{eq:film}.
Our initial methodology involves leveraging a pre-trained publicly available SSL model to obtain a sequence of self-supervised embeddings ($s_1,\dots,s_T$) where $T$ is the number of such vectors.
We use the last layer of SSL models for simplicity.
For CGAN optimization, we also explore using auxiliary speaker embedding network for deep feature loss~\cite{kataria2020feature,kataria2020analysis,kataria2021perceptual} to preserve speaker identity during BWE.
Self-FiLM can be seen as top-down conditioning, while DFL can be seen as bottom-up conditioning (of speaker identity).
Self-FiLM CGAN is thus a semi-supervised model, while the CycleGAN version is entirely unsupervised.
Our preliminary experiment on applying Self-FiLM to x-vector indicated over-fitting on in-domain data and hence, we defer it to future work.

\section{Experimental Setup}
\label{sec:exp}
As stated in Sec.~\ref{sec:intro}, our goal is to improve ASV performance on telephone test sets.
Richer wideband training data is available to train x-vector (LResNet, RawNet3) and SSL models.
X-vectors are trained on VoxCeleb~\cite{nagrani2020voxceleb}, narrowband VoxCeleb (\emph{VoxCeleb\_narrow}), and \ac{SRE} telephone data (\emph{SRE\_telephone}).
VoxCeleb (1\&2 combined) contains 2700+ hrs of audio from 7365 speakers in the wild.
\emph{VoxCeleb\_narrow} is created by removing upperband information (4-8KHz).
\emph{SRE\_telephone} is created by combining SRE Superset~\cite{sadjadi2021nist} and SRE16 eval data~\cite{kataria2022time} which includes Tagalog and Cantonese (YUE) languages.
For BWE training, VoxCeleb\_narrow and VoxCeleb are used for domains A and B.
We test on three sets that cover a variety of languages, acoustic environments: \emph{SRE16-YUE-eval40} (40\% speakers (40) from evaluation set of SRE16 Cantonese), \emph{SRE-CTS-superset-dev} (99 speakers from CMN (Mandarin) and YUE (Cantonese) languages), and \emph{SRE21-audio-eval}~\cite{kataria2022time}.
For details, reader can refer to \cite{kataria2022time}.
Note that the narrowband signals are also resampled from native 8KHz to 16KHz.
LResNet34 and RawNet3 are pre-trained on VoxCeleb+VoxCeleb\_narrow+SRE\_telephone with \ac{AAM} softmax (margin=0.3)~\cite{villalba2020advances} speaker classification loss.
Training configurations for GANs are obtained from \cite{kataria2022time}.
For ASV evaluation, we use \ac{EER} and \ac{minDCF} metrics with a target speaker prior probability of 0.05.
LResNet-\ac{PLDA} pipeline is used for scoring~\cite{kataria2022time}.
All models are trained with PyTorch on a 4x24GiB NVIDIA GPU workstation. 
The code will be available soon.

\begin{table}[htbp]
\centering
\caption{\label{tab:effect}
Investigating role of in-domain SSL model, $\alpha$, pre-extension, deep feature supervision loss in Self-FiLM CGAN.
}
\resizebox{0.48\textwidth}{!}{
\begin{tabular}{@{}lccc@{}}
\toprule
 \textbf{BWE type} & \textbf{SRE16-YUE} & \textbf{SRE-CTS} & \textbf{SRE21}\\
    \hline
    No BWE & 7.12 / 0.376 & 5.36 / 0.216 & 17.12 / 0.644 \\
    \hline
    \textit{\textbf{Training data for data2vec 2.0}} &&& \\
    Librispeech & 5.48 / 0.308 & \textbf{3.94} / \textbf{0.175} & 16.07 / 0.621 \\ 
    VoxCeleb & 5.52 / 0.305 & 3.98 / 0.176 & 15.09 / 0.604 \\ 
    SRE\_telephone & 5.40 / 0.307 & 3.97 / 0.176 & 15.45 / 0.608 \\ 
    SRE\_telephone+VoxCeleb8k & 5.56 / 0.308 & 3.99 / 0.177 & 14.81 / 0.598 \\ 
    \hline
    \textbf{VoxCeleb data2vec 2.0} & 5.52 / 0.305 & 3.98 / 0.176 & 15.09 / 0.604 \\
    + $\alpha$=0.5 & 5.59 / 0.316 & 4.02 / 0.178 & 14.59 / 0.595 \\ 
    + pre-extension & 5.43 / 0.303 & 3.95 / 0.176 & 15.57 / 0.612 \\ 
    + feature-domain DFL & 5.51 / 0.305 & 3.97 / 0.176 & 15.01 / 0.605 \\ 
    + time-domain DFL & 5.42 / \textbf{0.301} & 4.00 / 0.178 & \textbf{14.07} / 0.589 \\ 
    + $\alpha$=0.5+time DFL & 5.40 / 0.305 & 4.01 / 0.178 & 14.08 / \textbf{0.588} \\ 
    + $\alpha$=0.5+time DFL+pre-extend & \textbf{5.28} / 0.308 & 3.98 / 0.180 & 14.20 / 0.591 \\ 
\bottomrule
\end{tabular}
}
\vspace{-5mm}
\end{table}

\section{Results}
\label{sec:res}
\subsection{Exploring pooling methods and SSL models}
\label{sec:res_explore}
Here, we show the effectiveness of Self-FiLM under a wide variety of pooling methods and SSL model choice (Table~\ref{tab:explore}).
First row contains the baseline results without BWE.
With a CGAN BWE, we can see drastic improvement across all test sets and obtain a supervision loss value of 0.0048.
This recreates results from previous studies~\cite{kataria2022joint,kataria2022time} and establishes a strong baseline for further experiments as the CGAN has been tuned extensively.
Next, we investigate pooling methods using wav2vec 2.0 Self-FiLM.
Here, we see higher improvements with complex methods like mean+std, LDE, and ScaleAtt.
Note that BASE wav2vec 2.0 does not improve CGAN supervision loss.
Then, we investigate stronger SSL models with ScaleAtt pooling which all improves the baseline.
We note that the results do not necessarily improve consistently with stronger models as corroborated by prior studies~\cite{song2023exploring}.
This is perhaps due to a mismatch with the training data domain of the pre-trained SSL models. 
However, lower supervision loss and the observed improved GAN training reveal the potential of such models.

We visualize that Self-FiLM can extract discriminative information about speaker, language, and domain from SSL models (Fig.~\ref{fig:analysis}).
We apply t-SNE on FiLM activations from the first layer of CGAN generator.
In Fig.~\ref{fig:analysis}(a), we plot recordings of five random speakers and observe clustering by speaker identity.
In Fig.~\ref{fig:analysis}(b), we fix a speaker and discover clustering by the languages spoken.
We observed this for other speakers as well.
Finally, in Fig.~\ref{fig:analysis}(c), we observe clustering by two domains: CTS (conversational telephone) and AFV (wideband).

\subsection{Effect of in-domain training data for data2vec 2.0, FiLM strength ($\alpha$), pre-extension and deep feature loss}
\label{sec:res_effect}
We choose data2vec2.0 for further analysis (Table~\ref{tab:effect}) as it has the highest training efficiency~\cite{baevski2022data2vec}.
Test set names are shortened for brevity.
First, we train data2vec 2.0 on different datasets.
We find naturalistic wideband and narrowband data to be better than \ac{OOD} read speech corpus Librispeech.
We get even better results than Robust wav2vec 2.0 (Table~\ref{tab:explore}).
We combined VoxCeleb\_narrow and SRE\_telephone to observe great performance but it led to unstable GANs for further experiments with DFL.
This is perhaps because the existing training configuration of data2vec 2.0 is optimized for wideband data.
Using VoxCeleb data2vec 2.0, we conduct further experiments.
Using $\alpha=0.5$ on CGAN gives slight improvement, while \emph{pre-extension} gives slightly inconsistent gains.
With deep feature loss in CGAN training, we observe significant improvements, especially with temporal model (RawNet3), as CGAN operates in the time domain.
Finally, we try combinations of the above experiments.
We observe synergy in different test sets which can be advantageous for ASV \emph{fusion}~\cite{villalba2022advances20,villalba2022advances21}.

\subsection{Application to CycleGAN-based bandwidth extension}
\label{sec:res_cycle}
Here, we prove the compatibility of unsupervised BWE models based on CycleGAN with Self-FiLM.
In Table~\ref{tab:cycle}, we note the benefit of using (default) Librispeech and VoxCeleb data2vec2.0.
Performance on SRE21 is greatly improved while other test sets benefit from RawNet3 based DFL (in cycle and identity loss).
On SRE21, there is degradation in last row perhaps due to 1) usage of CGAN hyper-parameter configuration or 2) using identical RawNet3 for the other generator, which learns the reverse mapping.
In the future, we can utilize larger computational budget to 1) discover ideal CycleGAN hyper-parameters, and 2) explore different RawNet3 and data2vec 2.0 data configurations.
We can also explore joint training of CycleGAN and SSL models in the future.

\begin{table}[htbp]
\centering
\caption{\label{tab:cycle}
Integration of unsupervised BWE with Self-FiLM
}
\resizebox{0.46\textwidth}{!}{
\begin{tabular}{@{}lccc@{}}
\toprule
 \textbf{BWE type} & \textbf{SRE16-YUE} & \textbf{SRE-CTS} & \textbf{SRE21}\\
    \hline
    No BWE & 7.12 / 0.376 & 5.36 / 0.216 & 17.12 / 0.644 \\
    BWE w/o Self-FiLM & \textbf{4.95} / 0.294 & 3.99 / \textbf{0.176} & 17.58 / 0.681 \\ 
    Librispeech Self-FiLM & 4.97 / 0.297 & 4.16 / 0.183 & 15.96 / 0.637 \\ 
    VoxCeleb Self-FiLM & 5.47 / 0.317 & 4.52 / 0.197 & \textbf{14.02} / \textbf{0.609} \\ 
    + time-domain DFL  & 5.09 / \textbf{0.290} & \textbf{3.98} / 0.180 & 16.75 / 0.637 \\ 
\bottomrule
\end{tabular}
}
\vspace{-6mm}
\end{table}

\section{Conclusion}
We proposed \emph{Self-FiLM} to conditions a BWE model with the self-supervised representation of the input signal itself.
We demonstrate our approach's generality and contribute to the understanding of high-level information in SSL embeddings. 
We also reproduce and extend the claims of the Robust wav2vec 2.0 by training data2vec 2.0 on mixed-bandwidth in-domain data.
In our framework, we showed data2vec 2.0 is compatible with narrowband inputs, prior BWE (\emph{pre-extension}) model, and deep feature loss-based BWE.
Finally, we extend Self-FiLM to combine CycleGAN and data2vec 2.0 for completely unsupervised solution.
In the future, we can apply Self-FiLM to other speech applications and diffusion models.

\clearpage

\bibliographystyle{IEEEtran}
\bibliography{mybib}

\end{document}